\documentclass[aps, pra, showpacs, twocolumn, amsfonts, amsmath, amssymb, superscriptaddress, floatfix]{revtex4}
 
\usepackage[T1]{fontenc} 
\usepackage[latin9]{inputenc}
\usepackage{color}   
\usepackage{graphicx}
\usepackage{epstopdf} 
\usepackage{ulem}
 
\graphicspath{pics}

\def\be{\begin{equation}}  
\def\ee{\end{equation}} 
\def\bea{\begin{eqnarray}}  
\def\eea{\end{eqnarray}}

\begin{document}

\title{Quantum dark solitons in Bose gas confined in a hard wall box} 

\author{Andrzej Syrwid} 
\affiliation{
Instytut Fizyki imienia Mariana Smoluchowskiego, 
Uniwersytet Jagiello\'nski, ulica Profesora Stanis\l{}awa \L{}ojasiewicza 11 PL-30-348 Krak\'ow, Poland}
 
 
 
\author{Krzysztof Sacha} 
\affiliation{
Instytut Fizyki imienia Mariana Smoluchowskiego, 
Uniwersytet Jagiello\'nski, ulica Profesora Stanis\l{}awa \L{}ojasiewicza 11 PL-30-348 Krak\'ow, Poland}
\affiliation{Mark Kac Complex Systems Research Center, Uniwersytet Jagiello\'nski, ulica Profesora Stanis\l{}awa \L{}ojasiewicza 11 PL-30-348 Krak\'ow, Poland
}

\pacs{03.75.Lm, 03.75.Hh, 42.65.Tg} 

\begin{abstract}
Schr\"odinger equation for Bose gas with repulsive contact interactions in one-dimensional space may be solved analytically with the help of the Bethe ansatz if we impose periodic boundary conditions. It was shown that for such a system there exist many-body eigenstates directly corresponding to dark soliton solutions of the mean-field equation. The system is still integrable if one switches from the periodic boundary conditions to an infinite square well potential. The corresponding eigenstates were constructed by M. Gaudin. We analyze the weak interaction limit of Gaudin's solutions and identify parametrization of eigenstates strictly connected with single and multiple dark solitons. Numerical simulations of measurements of particle's positions reveal dark solitons in the weak interaction regime and their quantum nature in the presence of strong interactions.  
\end{abstract}

\maketitle

\section{Introduction}
\label{intro}

Self-reinforcing solitary solutions of non-linear wave equations maintaining their shape during time evolution are called solitons and appear in various fascinating phenomena \cite{KivsharOpticalSol}. Solitons are particularly investigated in non-linear optics and ultra-cold atomic gases. In the latter case bosons may form a Bose-Einstein condensate (BEC) where all atoms occupy the same single-particle state and the many-body wave function factorizes into the product of identical single-particle states \cite{pethicksmith}. In the presence of inter-particle interactions the product state describes atoms living in an averaged potential coming from the  milieu of other identical particles (mean-field description). The behaviour of the single-particle state is determined by the Gross-Pitaevskii equation (GPE) which has analytical bright and dark soliton solutions in one-dimensional (1D) space for attractively and repulsively interacting system, respectively \cite{pethicksmith}. GPE gives very accurate description of the solitons realized experimentally so far \cite{KivsharOpticalSol, burger1999, denschlag2000, strecker2002, khaykovich2002, becker, Stellmer2008,  Weller2008, Theocharis2010}. 
 
Theoretical investigations of quantum nature of dark solitons, i.e. properties which go beyond the standard Bogoliubov corrections \cite{dziarmaga2004,Mishmash2009_1,Mishmash2009_2,Dziarmaga2010,Mishmash2010,delande2014,kronke15,Hans2015},  employ the phase imprinting method \cite{burger1999,denschlag2000,becker,Stellmer2008,Yefsah}--- starting with the system in the ground state one can carve a soliton by shining a short laser pulse on a half of the atomic cloud. Sufficiently long time evolution reveals quantum character of the dark soliton when the soliton position  fluctuates on a length scale which can be much greater than the soliton width. Hence, the soliton location has to be considered as a quantum degree of freedom described by a probability density whose standard deviation increases with time \cite{dziarmaga2004,delande2014}. It is not easy to observe the many-body effects experimentally because it requires relatively small number of particles in the system in order to reduce atomic losses that are able to kill the quantum character of the solitons \cite{roberts2000,plodzien2012}. In the field of ultra-cold atomic gases the experimental techniques evolve very rapidly giving an opportunity to investigate systems for which the many-body effects play a key role \cite{Theocharis2010,dziarmaga2004, Lai89, Lai89a, castinleshouches, Weiss09, delande2013, corney97, corney01, martin2010b, delande2014, kronke15, Mishmash2009_1, Mishmash2009_2, Hans2015,Boisse2017}. 
 
The Schr\"odinger equation for some quantum many-body systems possesses analytical solutions. Generally, it happens for systems in lower dimensions when one can use a brilliant concept of the Bethe ansatz \cite{Bethe31}. It turns out that the Bethe ansatz approach is crucial in the analysis of bosonic and fermionic 1D systems interacting via point-like $\delta$-potentials \cite{Korepin93, Gaudin, Oelkers2006}. Non-relativistic ultra-cold bosons in 1D space with contact interactions are described by renowned Lieb-Liniger model \cite{Lieb63,Lieb63a}. One can expect dark soliton solutions in the presence of repulsive inter-particle interactions. In contrast to attractive interactions for which bright soliton solution corresponds to the lowest energy state, dark solitons have to be a reflection of excitations. Furthermore, if the system satisfies periodic boundary conditions all energy eigenstates are also eigenstates of the unitary operator that translates all particles by the same distance. Hence, they satisfy translation symmetry which is broken in the case of the mean-field soliton solutions. For those reasons the identification of dark soliton-like many-body eigenstates was particularly difficult. The conjecture and various evidences that the eigenstates belonging to the so-called type II excitation spectrum of the Lieb-Liniger model are strongly connected with the mean-field soliton solutions can be found in the literature \cite{kulish76, ishikawa80, komineas02, jackson02, kanamoto08, kanamoto10, karpiuk12, karpiuk15, sato12, sato12a, sato16, Gawryluk2017}. Especially, it has been shown that dark soliton signatures are visible in the reduced single-particle density if the system is prepared in a proper superposition of the eigenstates coming from the second branch of the excitation spectrum \cite{sato12, sato12a, sato16}. Ultimately, it was demonstrated that dark soliton signatures emerge in the course of measurement of particle positions if the system is prepared initially in a type II eigenstate \cite{ Syrwid2015, Syrwid2016}.  

In the presence of hard-wall boundary conditions the system does not possess the translation symmetry but the question if there are many-body eigenstates directly corresponding to mean-field dark soliton solutions in this case is still open. It is clear that in the limit of infinitely weak repulsion, one can identify many-body eigenstates that reveal density notches present in plots of the single-particle probability densities. Genuine solitons may appear if the inter-particle interactions are turned on and solutions of the GPE reveal phase flips and density notches whose widths are much smaller than the size of the hard-wall box. The goal of the present work is to show that there are many-body eigenstates that correspond directly to the solitonic structures.  We will show that the phase flips and signatures of the density notches can be observed for arbitrary strength of the repulsive particle interaction. Therefore, in order to simplify the nomenclature in the present manuscript this class of states we will be sometimes dubbed {\it dark solitons} even in the nearly non-interacting and strongly interacting cases. 

\section{Lieb-Liniger model}
\label{liebliniger}

Ultra-cold system consisting of $N$ bosons with repulsive contact interactions in the 1D space may be described by the following Lieb-Liniger Hamiltonian \cite{Lieb63}
\be
H=\int\limits_0^L\mathrm{d}x\left[\partial_x\hat\psi^\dagger\partial_x\hat\psi+c\hat\psi^\dagger\hat\psi^\dagger\hat\psi\hat\psi\right],
\label{h}
\ee
where $\hat\psi$ is the canonical Bose field operator and $L$ is the system size. The units have been chosen so that $2m=\hbar=1$ where $m$ is the particle mass. The dimensionless parameter

\be
\gamma=\frac{c}{n},
\ee
reflects the strength of the interactions. The coupling constant $c>0$ and $n=\frac{N}{L}$ denotes the average density of particles. One deals with the weak interaction limit when $\gamma\ll1$ and strongly interacting impenetrable bosons for $\gamma\gg 1$ \cite{Lieb63,Lieb63a}.

\subsection{Gaudin's solution in the presence of hard walls}

The Hamiltonian (\ref{h}) with periodic boundary conditions has analytical solutions that can be found with the  help of the Bethe ansatz \cite{Korepin93}. The Lieb-Liniger eigenstates are parametrized by a collection $\{k\}$ of $N$ real (if $c>0$) numbers called quasimomenta \cite{Gaudin}. It is clear that the integrability of the model can be easily broken by switching on a trapping potential. However, the model is still integrable in the presence of hard-wall boundary conditions. The solutions were constructed by M. Gaudin \cite{Gaudin71, Gaudin, Batchelor05, Tomchenko17}.
Assuming that $0\leq x_1 \leq x_2 \leq \ldots \leq x_N \leq L$ we have to fulfil two conditions 
\be
\Psi(x_1 \! = 0, x_2, \ldots, x_N) =\Psi(x_1, x_2, \ldots, x_N \! = \! L) = 0.
\label{conds}
\ee

Firstly, using McGuire's optical analogy \cite{McGuire}, one constructs solutions in the semi-infinite 1D space $x_i \geq 0$ vanishing at $x_1=0$. The solutions turn out to be superpositions of elementary waves  and take the following form
\begin{flushleft}
$
\displaystyle{\Psi(\{x\},\{q\}) =\sum_{\sigma \in \mathcal{S}_{N}} \sum_{\{\epsilon\}}\epsilon_1 \epsilon_2 \cdot \cdot \cdot \epsilon_N\exp\left[ i \sum_{s=1}^{N} q_{\sigma(s)} x_s\right]}
$\end{flushleft}
\be
\times \prod_{i<j}\left( 1-\frac{i c}{q_{i}+q_{j}} \right)\left( 1+\frac{i c}{q_{\sigma(i)}-q_{\sigma(j)}} \right),
\label{GauSol}
\ee
where $\epsilon_i = \pm 1$, $q_{i}=\epsilon_{i}|k_{i}|$ (with $0<|k_1|<\ldots < |k_N|$). The first sum has to be taken over all possible $N$-element combinations of $\pm1$. The  second sum refers to all permutations $\sigma$ from the permutation group $\mathcal{S}_N$. Therefore, calculation of a single value of (\ref{GauSol}) requires summation over $2^N N!$ elements \cite{Gaudin71, Gaudin, Batchelor05, Tomchenko17}.

Secondly, imposing the vanishing of the wave function at $x_N=L$ one obtains elegant system of coupled equations
\be
L k_i = \pi n_i +\sum_{ \substack{\, j=1 \\ j\neq i } }^{N}\left[ \mathrm{arctan} \frac{c}{k_i - k_j} +  \mathrm{arctan}  \frac{c}{k_i + k_j}  \right],
\label{GauEqs}   	
\ee
called Gaudin's equations where integer numbers $n_i$ are parameters that define an eigenstate. In the case of repulsive inter-particle interactions the system of Gaudin's equations (\ref{GauEqs}) has unique real solutions $k_i$ \cite{Tomchenko17}.
In order to deal with admissible physically different solutions one considers only $k_{i=1,\ldots , N}>0$ and the set of integers $\{n \}$ where $1\leq n_1\leq n_2 \leq \ldots \leq n_N$ \cite{Gaudin71, Gaudin, Batchelor05, Tomchenko17}. 
By construction the set $\{k\}$ determines the energy of the eigenstate (\ref{GauSol}) \cite{Gaudin, Gaudin71, Batchelor05} 
\vspace{-0.0cm}
\be
E_{\{k\}}=\sum_{j=1}^{N}k_j^2.
\label{GauEnergy}
\ee
We would like to stress that even if two or more $n_i$ parameters are equal, the resulting solutions $k_i$ of (5) are always distinct.

\subsection{Collective soliton-like excitations}

For the non-interacting system ($c\rightarrow 0^+$) the solutions (\ref{GauSol}) reduce to superposition of products of sine functions (see Appendix)
\be
\Psi(\{x\},\{k\})\stackrel{c\rightarrow 0^+}{\propto}\sum_{\sigma\in \mathcal{S}_N}\prod_{s=1}^{n}\sin \left(k_{\sigma(s)} x_s\right),
\label{GauReduced} 
\ee
with $k_j \stackrel{c\rightarrow 0^+}{\longrightarrow} \pi n_j /L $ for $n_j= 1, 2, 3, \ldots$ (negative values are physically equivalent to non-negative ones). Obviously, they coincide with the well known solutions of the problem of non-interacting particles in an infinite square well potential. If we choose identical integer parameters $n_i=j$ in the Gaudin's equation (\ref{GauEqs}), all elements of the set $\{k\}$ approach  $\pi j /L$ in the limit $c\rightarrow 0^+$ but they are always slightly different if $c$ is not strictly equal to 0. 

In the non-interacting case of Bose gas confined in the 1D box of length $L$ the solutions resembling solitons correspond to product states where all particles occupy the same excited eigenstate of the single-particle problem,
\be
\Phi_{sol}^{j-1}(\{x\}_N)\propto \prod_{s=1}^{N}\sin\left( \frac{\pi j x_s}{L}\right),
\label{SolitonSol} 
\ee
where $j=2,3,\ldots$. It is clear that the wave function (\ref{SolitonSol}) reveals $j-1$ density notches and phase flips and thus resembles signatures of $j-1$ dark solitons. When $j=1$ it reproduces the ground state. In the $c\rightarrow 0^+$ limit the eigenstates (\ref{GauSol}) reduce to (\ref{SolitonSol}) for the following collections of the integer numbers in (\ref{GauEqs})
\be
\mathcal{T}(j): n_{1}=n_{2}=\ldots =n_{N}=j.
\label{SolitonCol} 
\ee
We believe that in the presence of particle interactions there is a range of the system parameters where the eigenstates (4), with quasimomenta $\{k\}$ parametrized by (9), not only resemble but reveal dark solitons unambiguously.

Looking at the expression (\ref{SolitonCol}) we notice that $\mathcal{T}(1+s)$ corresponds directly to $s$-fold collective excitation of the aforementioned non-interacting system. For convenience, we will use this terminology also in the presence of inter-particle repulsion. The excitation given by $\mathcal{T}(j)$ will be dubbed as \textsl{the first collective excitation} for $j=2$ and \textsl{the higher ($j-1$)th collective excitation} for $j>2$.

\subsection{Numerical method}

Properties of the many-body eigenstates $\Psi(\{x\},\{q\})$ are hidden in the structure of terms in (\ref{GauSol}) whose number dramatically grows with increasing number of particles $N$. Hence, despite the fact that one has the analytical solutions it is not easy to understand their features. In order to study the eigenstates of the many-body system we will simulate measurements of particle positions. 

In order to simulate results of the measurements one has to choose randomly positions of $N$ particles according to the $N$-dimensional probability density. In practice it is often done sequentially: one by one particle at each step calculating conditional probability density for a choice of a next particle \cite{javanainen96, dziarmaga03, dziarmaga06, Syrwid2015, Dagnino09, Kasevich20016}. This procedure is very difficult in the present case. 
Instead, we follow an equivalent idea of usage of the Monte Carlo algorithm of Metropolis {\it et al.} \cite{Metropolis1953}. As in Refs.~\cite{Gajda_PauliCrystal,Syrwid2016} we perform Markovian walk in the configuration space. That is, by sampling  $N$-dimensional probability distribution $\left|\Psi(x_1,\ldots,x_N,\{q\})\right|^2$ we generate collections of sets $\mathcal{X}=$ $\{x_1,\dots,x_N\}$ of particles' positions. The procedure is based on step by step acceptance of sets $\mathcal{X}'$  with probability  $p={\rm min}\left(1,|\Psi(\mathcal{X}')|^2/|\Psi(\mathcal{X})|^2\right)$, where $\mathcal{X}$ is the previously accepted set of particles' positions.  Although the method is very efficient, only a few body systems are numerically attainable. In the studies of the Lieb-Liniger model with periodic boundary conditions we were able to investigate 8-particle system \cite{Syrwid2015, Syrwid2016}. There, number of terms in the Bethe ansatz solutions increases like $N!$ with an increase of the total number of particles $N$. In the present case  we need to use the Gaudin's solution (\ref{GauSol}) where the number of terms proliferates $2^N$ times faster than in the previously used Bethe ansatz solutions. Hence, in order to collect reliable statistics, we perform simulations for the system consisting of $N=6$ and 7 bosons only.

\section{The analysis of collectively excited many-body eigenstates} 
\label{weak}
   
The ground and collectively excited states of  a Bose system are usually described  with the help of the mean-field approximation if we deal with the weak interaction regime ($\gamma\ll 1$). The assumption that all particles occupy the same single-particle state leads to GPE \cite{pethicksmith}. Analytical solutions of GPE in the presence of the hard wall box potential exist and are given by Jacobi elliptic functions \cite{Carr_HWSoliton}. If $L\gg\xi$, where $\xi=1/\sqrt{cn}$ is the so-called healing length \cite{pethicksmith}, the analytical solution related to a dark soliton can be approximated by the following function \cite{dziarmaga2004}  
\be
 \psi(x)  =  \left\{ \begin{array}{ll}
\!\!  -\sqrt{n}\tanh\left(\frac{x}{\xi}\right) & \textrm{for $0\le x\ll L/2-\xi$}\\
\!\!  \sqrt{n}\tanh\left(\frac{x-L/2}{\xi}\right) & \textrm{for $\xi\ll x\ll L-\xi$}\\
\!\!   \sqrt{n}\tanh\left(\frac{L-x}{\xi}\right) & \textrm{for $L/2+\xi\ll x\le L$.}
\end{array} \! \! \!  \right. 
\label{tanh}
\ee
The healing length $\xi$ describes a typical distance over which the condensate wave function forgets about, e.g.,  boundaries of the system. Therefore, in a finite system with hard walls positioned at the boundaries we expect disturbance of the particle density close to the boundaries on the same length scale as the width of the dark soliton notch.   

When quantum many-body effects are taken into account, one may expect that the position of the dark soliton starts fluctuating in different realizations of the particle positions measurements. The fluctuations are the more significant the stronger interactions are present \cite{Syrwid2015,Syrwid2016}. Hence, we expect that the averaged particle density $\rho(x)$, i.e. averaged over many realizations of the measurement process, should reveal the shallower dark soliton notches the stronger repulsion is.   

\subsection{First collective excitation}
\label{ssols}

The signatures of single dark soliton are expected  to be observed in the weak interaction limit when  one prepares the system in the eigenstate (\ref{GauSol}) choosing the parameters in (\ref{GauEqs}) so that $\forall_{i=1,\ldots N}: n_{i} = 2$.  We have performed numerical simulations of the particles positions measurement for 6- and 7-body systems confined in a box of length  $L=1$. Single realization of the detection process produces a sequence of $N$ positions only. Therefore, in order to investigate averaged particle density, we have repeated the measurement procedure starting with the same many-body eigenstate many times. Collecting all results of the simulations in a histogram allows us to look at the average particle density. 
\begin{figure}[h] 	        
\includegraphics[width=1.\columnwidth]{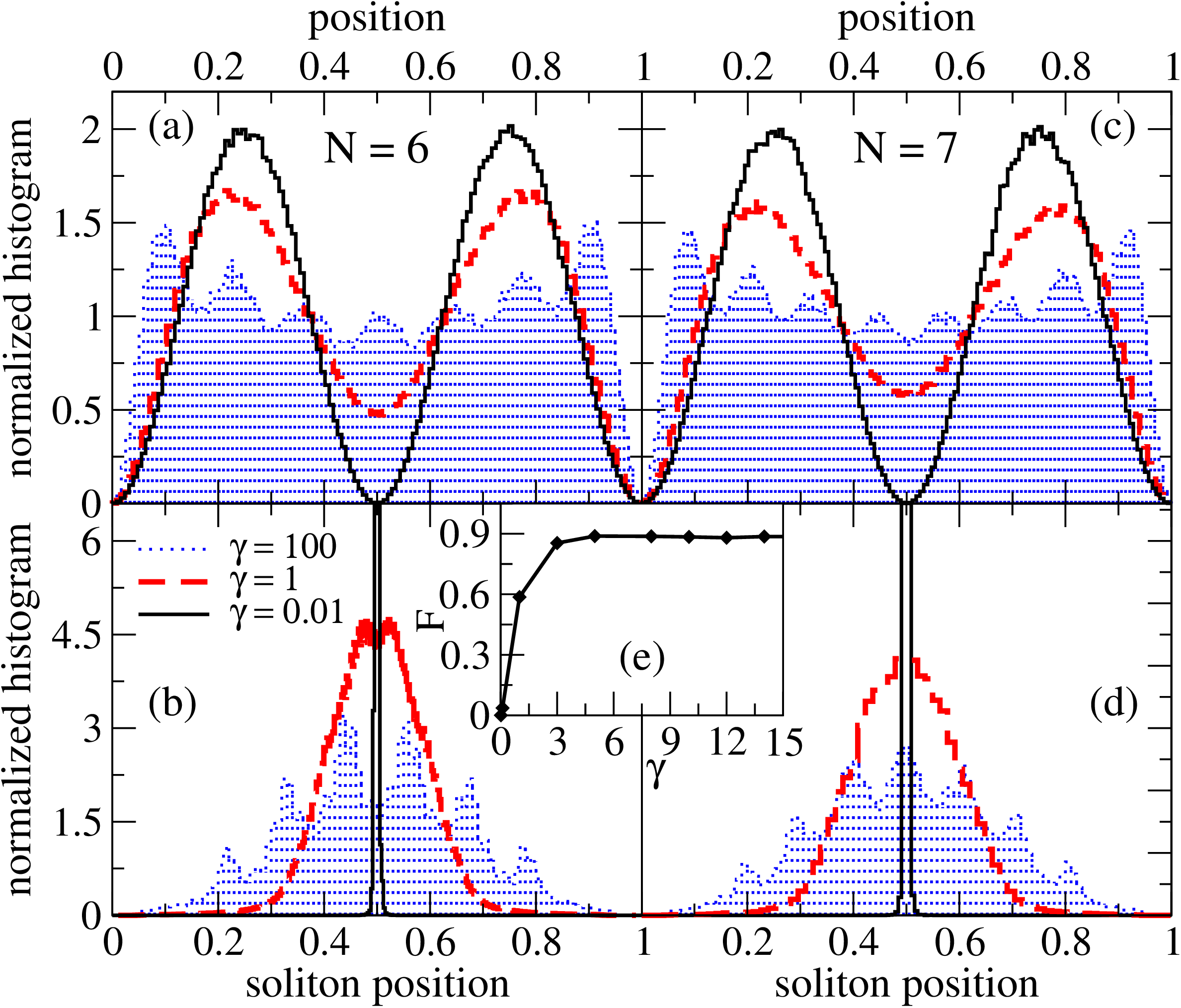}       
\caption{(color online) Single dark soliton:  top panels present the averaged particle densities for 3 different interaction strengths $\gamma=0.01$ (solid lines), $1$ (dashed lines) and 100 (dotted filling) and for $N=6$ (a) and $N=7$ (c).  In the case of the weak interaction regime ($\gamma=0.01$) the averaged particle density reproduces the non-interacting solution described by (\ref{SolitonSol}) with $j=2$. In the presence of intermediate and strong interactions ($\gamma = 1$ and 100) one observes smearing of the density notch clearly visible at the center of the box for $\gamma=0.01$ and $\gamma=1$. The effect is caused by fluctuations of the dark soliton position in different realizations of the measurement process, see discussion in the main text. Distributions of the soliton positions, identified with the positions of the phase flips of the wave function (\ref{lastwavefun}), are presented in bottom panels, $N=6$ (b) and $N=7$ (d).  The degree of filling of the density notch in averaged particle density,  Eq.~(\ref{SolitonDepth}), obtained for different interaction strengths in the case of the 7-body system is shown in (e). We see that $\mathrm{F}$ saturates for $\gamma\approx5$ at the value $\mathrm{F} \approx 0.9$. The system was prepared in the eigenstate (\ref{GauSol}) parametrized by the set $\{n\}$ satisfying (\ref{SolitonCol}) with $j=2$. The size of the box is $L=1$. }
\label{f1}
\end{figure} 

In panels (a) and (c) of Fig.~\ref{f1} we present histograms of the obtained particle positions in many realizations of the detection process for $N=6$ and 7 and for a wide range of the interaction strength ($\gamma =0.01, 1, 100$). Smearing of the density notch with increasing $\gamma$ is clearly visible at the center of the box. It is caused by the fact that the soliton position varies randomly from realization to realization and the range of the fluctuations is the larger the stronger interactions are \cite{Syrwid2015,Syrwid2016}. 
In a single realization of the measurement process one obtains a set of particle positions $\{x_1,\dots,x_N\}$. Let us choose from this set arbitrary $N-1$ positions and consider the eigenstate (\ref{GauSol}),  parametrized by the set of $n_i$ that satisfy (\ref{SolitonCol}) with $j=2$, as a single-particle wave function of the last remaining particle $x_i=x$, 
\be 
\psi(x)\propto\Psi(x_1,\dots,x_{i-1},x,x_{i+1}\dots,x_N).
\label{lastwavefun}
\ee
It turns out that the wave function $\psi(x)$ reveals density notch and a phase flip at the center of the notch.  We identify the position of the phase flip with the position of the dark soliton. 
The distributions of the phase flip positions in the case of different interaction strengths are depicted in Fig.~\ref{f1} for $N=6$ (b) and $N=7$ (d). Additionally, in Fig.~\ref{f1}(e), we present the quantity
\be
\mathrm{F} \equiv  \rho(L/2), 
\label{SolitonDepth} 
\ee
which measures the degree of filling of the notch observed in the averaged particle densities $\rho(x)$ versus $\gamma$ parameter in the case of $N=7$. The degree of filling $\mathrm{F}$ saturates already for $\gamma \approx 5$. For strong interactions ($\gamma = 100$) one can also observe $N+1$ oscillations in the histograms, cf. panels (a) and (c) of Fig.~\ref{f1}. In the ground state case and for a small particle number one can expect $N$ local maxima in the average particle density that correspond to average locations of $N$ bosons with strong repulsive interactions --- the density should be identical to the density of non-interacting fermions when $\gamma\rightarrow\infty$ (Tonks-Girardeau limit) \cite{girardeau1960, Paredes2004}. In Fig.~\ref{f1} we see one local maximum more which seems to be related to the fact that if there is a phase flip between neighbouring particles, then their relative distance is modified and on average it results in an additional oscillation in the density profile. We expect that the oscillations can be visible for small $N$ only and for large particle number they will become negligible, i.e. the averaged  particle density profile will be almost flat (except the edges of the system).

The distributions of positions of the phase flip in the case of strong repulsion ($\gamma = 100$) also reveal oscillations [dotted histograms Fig.~\ref{f1}(b) and (d)]. We note that the positions of local maxima of such distributions roughly coincide with positions of local minima of corresponding averaged particle densities (dotted histograms in Fig.~\ref{f1}(a) and (b), respectively). Moreover, dealing with even (odd) number of particles we observe that in the presence of strong inter-particle interactions the distribution of the phase flip positions has the minimum (maximum) in the center of the box $x=L/2$. The presence of the oscillations and the correlations between positions of the maxima in the averaged particle densities and positions of the minima in the distributions of phase flip positions results from the fact that for large $\gamma$ at a space point where there is greater probability to observe a particle, there must be smaller probability to find the phase flip. 

\begin{figure}[h] 	            
\includegraphics[width=1.\columnwidth]{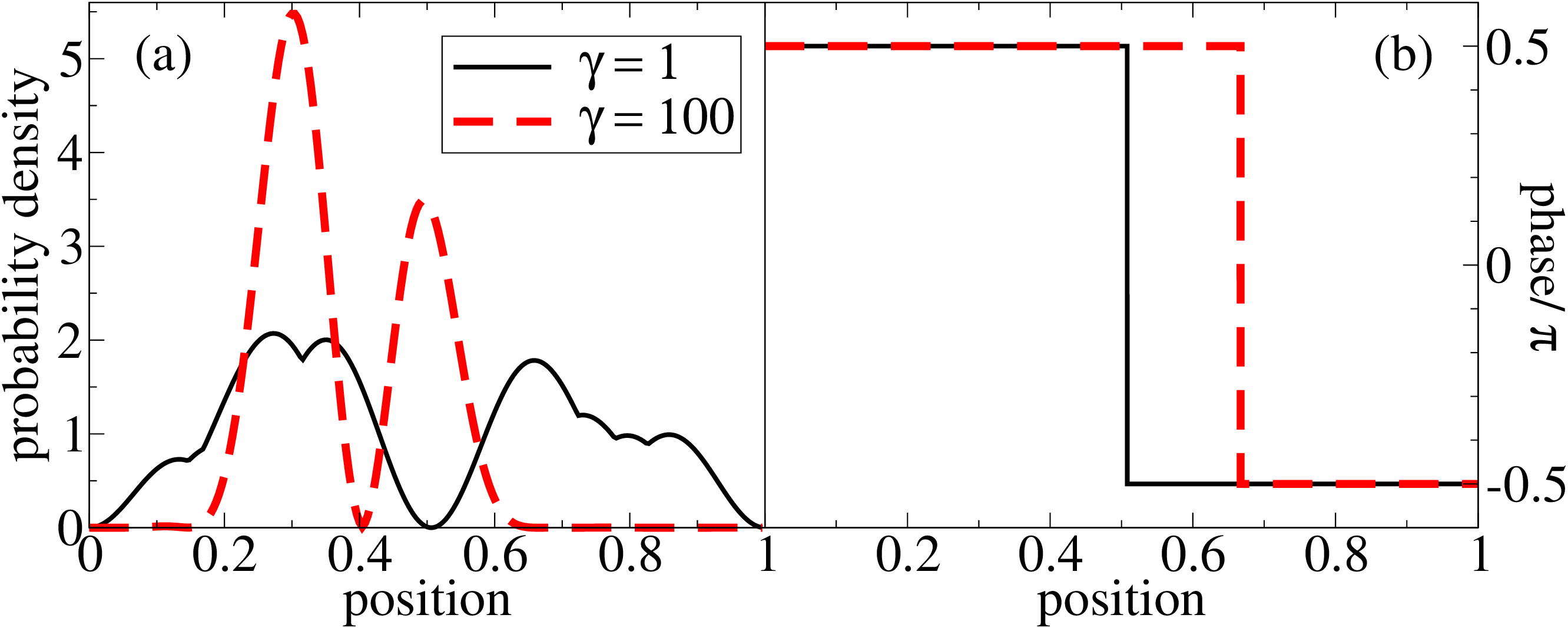}         
\caption{(color online) Single realizations of the detection process in the case of intermediate and strong interactions  for eigenstates (\ref{GauSol}) parametrized by (\ref{SolitonCol}) with $j=2$. Panel (a): typical probability densities of the wave function (\ref{lastwavefun}) for the last particle in 7-particle system for $\gamma=1$ (solid black line) and $\gamma=100$ (dashed red line). In (b) the corresponding plots of the phase of the last particle wave function (\ref{lastwavefun}) are presented. The size of the box is $L=1$.}
\label{f2}   
\end{figure}  

The significant quantum many-body effects appear for $\gamma>0.1$ but they do not destroy the signatures of solitons like the density notch and phase flip in single realizations of the detection process. Especially the phase flip can be clearly observed even in the very strong interaction regime ($\gamma \gg 1$). In Fig.~\ref{f2} we show probability density and phase of the wave function (\ref{lastwavefun}) for the last 7th particle (provided 6 particles have been already detected) in the 7-particle system obtained in a single realization of the measurement for $\gamma=1$ and $100$.
One easily notices that every measurement leaves its mark on the wave function. In the Fig.~\ref{f2}(a) we observe 6 slight incisions in the profile of the probability density for $\gamma=1$. The stronger inter-particle repulsion is, the deeper incisions are observed. In the case of strong repulsion ($\gamma=100$) we note that the probability density is essentially nonzero only in regions away from the measured positions of particles. The reason is the strong repulsion does not allow for detection of two particles close to each other. Therefore, it is very difficult to establish where is the phase flip of the wave function (\ref{lastwavefun}) by looking at the profile of the probability density only  if $\gamma\gg 1$ --- compare Fig.~\ref{f2}(a) and Fig.~\ref{f2}(b).

 In the non-interacting case the many-body eigenstate that we analyse here reduces to a simple product state (\ref{SolitonSol}) with $j=2$. Then, the parity symmetry, which is fulfilled by the Hamiltonian (\ref{h}), becomes apparent because such a state is also an eigenstate of the operator that transforms each $x_i$ in the following way $(x_i-L/2)\rightarrow -(x_i-L/2)$. The resulting average particle density vanishes at the centre of the box, cf. Fig.~\ref{f1}. Average particle densities related to $\gamma=1$ and $\gamma=100$ do not possess this property but it is not in contradiction with the parity symmetry. Indeed, the interactions between particles introduce coupling between states where all particles occupy anti-symmetric modes with states where some even number of particles occupy symmetric modes --- superposition of such states is also an eigenstate of the parity operator.
  
\subsection{Higher collective excitations}
\label{msols}

\begin{figure}[h] 	            
\includegraphics[width=1.\columnwidth]{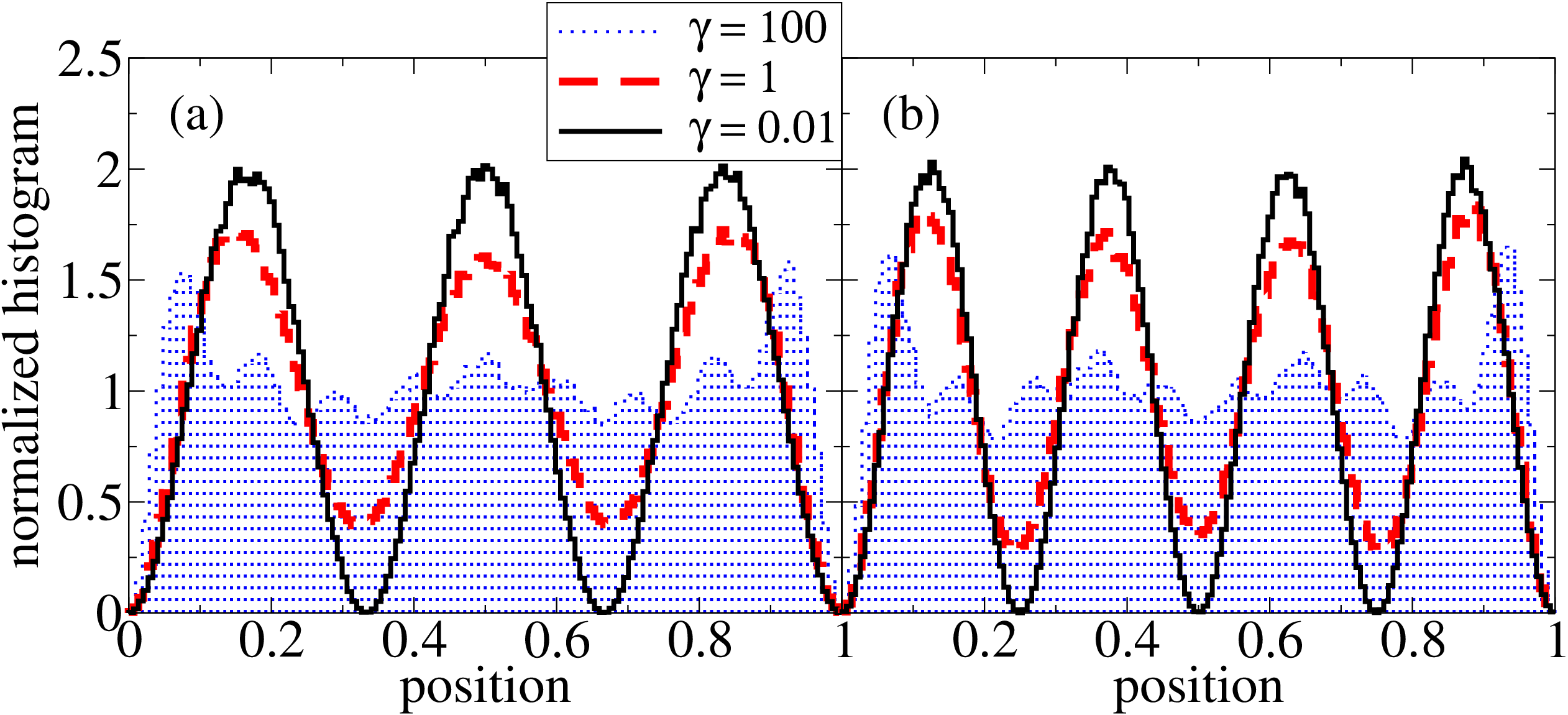}       
\caption{(color online) Multiple dark solitons: averaged particle densities for $N=7$ and for different interaction regimes as indicated by different values of $\gamma$. Panel (a) is related to the eigenstate (\ref{GauSol}) parametrized by (\ref{SolitonCol}) with $j=3$ while panel (b) corresponds to $j=4$ in (\ref{SolitonCol}). The densities reveal $j-1$ notches for the weak and intermediate interaction strength and small oscillations for strong interactions. The size of the box is $L=1$.	}
\label{f3}   
\end{figure}

Increasing the number $j$ in the eigenstate parametrization (\ref{SolitonCol}) one expects to observe increasing number of notches that appear in averaged particle densities for weak and intermediate interaction strength. We performed numerical simulations of particles detection for $N=7$ in the system prepared initially in the eigenstates parametrized by $n_i$ numbers  that fulfil (\ref{SolitonCol}) with $j=3$ and $j=4$. 
The results confirm that there are $j-1$ notches in the average probability densities for $\gamma=0.01$ and 1, see Fig.~\ref{f3}.  Analysis of single measurement realizations reveals $j-1$ phase flips in the wave function (\ref{lastwavefun}) for the last particle in the system. The phase flips are key signatures of dark solitons for the intermediate interactions and indicates where the solitons are localized in single realizations. The phase flips are also present in the strong interaction regime. 
  
 \vspace{-0.23cm}
 \vspace{-0.0cm}
\begin{figure}[h] 	             
\includegraphics[width=0.7\columnwidth]{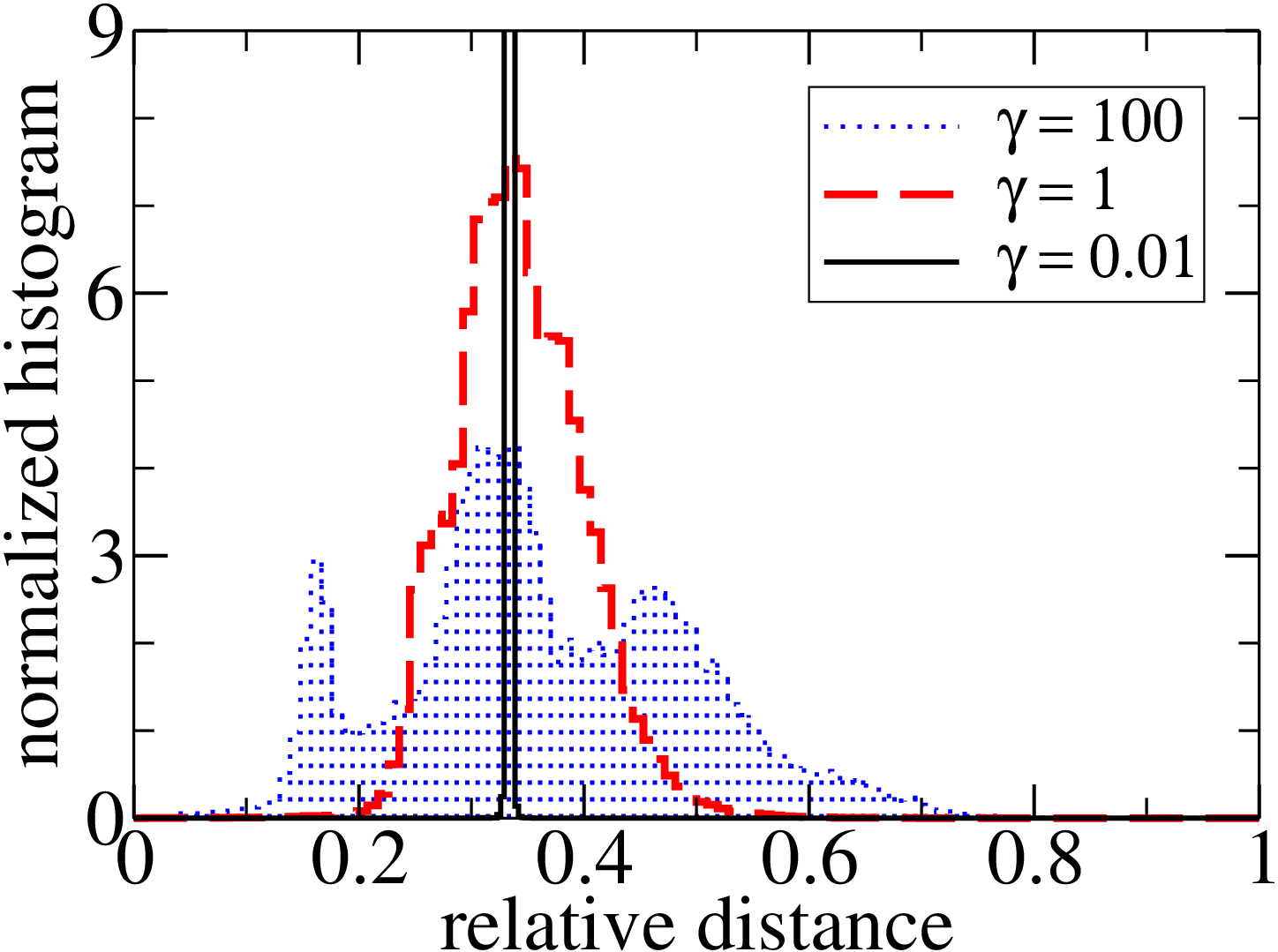}             
\caption{(color online) Distributions of relative distance between two phase flips in many realizations of the  measurement process  in the presence of three different interaction strengths: $\gamma=0.01$ (solid line), $\gamma=1$ (dashed line) and $\gamma=100$ (dotted filling). The system  of $N=7$ particles was initially prepared in the eigenstate (\ref{GauSol}) parametrized by (\ref{SolitonCol}) with $j=3$ and closed in the box of size $L=1$. } 
\label{f4}   
\end{figure} 

Similarly to the case of the first collective excitation, considered in the previous subsection, in the presence of intermediate and strong inter-particle repulsion the density notches are blurred because of fluctuations of positions of the phase flips. We also observe oscillations in the averaged particle densities for $\gamma=100$, see Fig.~\ref{f3}. One notices that the number of the oscillations is equal to $N+j-1$, i.e. the number of particles plus the number of phase flips.  In the strong interaction case and for small $N$ the average particle density related to the ground state would reveal $N$ local maxima. The presence of a phase flip between a pair of particles modifies its relative distance. It turns out that the presence of $j-1$ phase flips between different neighbouring particles results in $j-1$ additional local maxima in the average particle density as compared to the ground state case.

In the case of the eigenstate  parametrized by $n_i$ numbers that fulfill  (\ref{SolitonCol}) with $j=3$, the
distribution of distances between positions of two phase flips is obviously concentrated at $L/3\approx 0.33$ in the non-interacting case, cf. Fig.~\ref{f4}. For weak and intermediate interactions it is still localized around $L/3$ but its width increases with $\gamma$. This can be explained by the observation that the wave function (\ref{lastwavefun}) must drops at the edges on a length scale similar to half of the width of the density notches. If we now imagine that we merge the edges of the box, the shape of the modulus squared of the wave function (\ref{lastwavefun}) resembles not 2 but 3 solitons on a ring. Thus the expected mean separation between solitons should be approximately equal to $L/3$ but due to fluctuations of the soliton positions the width of the distribution increases with $\gamma$.

For $\gamma\rightarrow\infty$ we approach the Tonks-Girardeau regime where impenetrable bosons tend to localize at $L/N$ distances one from each other. Then, the phase flips are expected to be located half the way between two neighboring particles. Thus, the distributions of distances between two phase flips may reveal peaks at integer multiple of $L/N$. In Fig.~\ref{f4} we show the distribution for $\gamma=100$ where the peaks structure emerges. Similar structures were also observed in the case of periodic boundary conditions \cite{Syrwid2016}.

\section{Conclusions}
\label{conclusions}

We have considered  gas of bosons interacting via point-like $\delta$-potential confined in a one-dimensional hard wall box. Eigenstates of the system were analytically constructed by M. Gaudin and they were parametrized by a set of positive integers $\{n\}$. We show that Gaudin's solutions, in the limit of infinitely weak interactions, factorize to simple eigenstates of non-interacting particles in the box. In this case each number $n_i$  corresponds simply to a number of an excited eigenstate of a single-particle in the  square well potential. If all  numbers $n_i$ are equal to the same integer $j>1$, the average particle density for infinitely weak interactions resemble $j-1$ dark soliton-like notches. 

Genuine dark solitons may appear only for non-vanishing inter-particle interactions. In this case the numbers $\{n\}$ do not have clear interpretation and should be treated just as parameters defining a many-body eigenstate uniquely. Nevertheless, the eigenstates parametrized by $\forall_i$~$n_i = j$ do correspond to $j-1$ dark solitons. In order to show it we have performed numerical simulations of measurements of particle positions. It turns out that the wave function before the measurement of the last particle always possesses dark soliton signatures. That is, there  are $j-1$ phase flips of the wave function and $j-1$ probability density notches for the weak and intermediate interactions. The small numbers of particles ($N=6,7$) we consider do not allow us to fulfil the conditions, $\gamma\ll 1$ and $\xi\ll L$. Hence, shape of the density notches do not reproduce a hyperbolic tangent function like in Eq.~(\ref{tanh}).
For strong interactions, the positions of the phase flips fluctuate strongly from one realization of the measurement process to another one that results in smearing of the notches in the average particle density.

We have also investigated the relative distance between neighboring phase flips in the case of $j=3$. The results show that depending on interaction strength there are specific distances the phase flips localize more eagerly.

\section*{Acknowledgments}
The authors express their sincere gratitude to D. Delande for  fruitful discussions. 
Support of the National Science Centre, Poland via projects: No.2016/21/B/ST2/01095 (A.S.) and No.2015/19/B/ST2/01028 (K.S.) is acknowledged.
 A.S. acknowledges support in the form of a special scholarship from the Marian Smoluchowski Scientific Consortium "Matter-Energy-Future", from the KNOW funding. This work was performed with the support of EU via Horizon2020 FET project QUIC (nr. 641122)
\color{black}

\section*{Appendix}

Let us consider the Gaudin's equations for 2-particle system with identical numbers $n_1 = n_2$ in (\ref{GauEqs}). Subtracting the particular equations one obtains
   \be
L\delta k =  2 \, \mathrm{arctan} \frac{c}{\delta k}, 
\label{app1}
\ee
where $\delta k = k_2 - k_1$. Taking the tangent of both sides and using the series expansion,  $\tan(x)= x +{\cal O}(x^3)$,
one notices that in the limit $c\rightarrow 0^+$
   \be
\frac{L}{2}(\delta k)^2 \approx c \Longrightarrow \lim_{ \, \, \, c\rightarrow 0^+} \frac{c}{\delta k} = 0. 
\label{app3behavior}
\ee
Last result holds true also in the case of $N$-particle system parametrized by $s\leq N$ identical numbers $n_j$. It obviously coincides with the statement that physically relevant solutions may always be ordered such that $0<k_1<k_2<\ldots <k_N$ \cite{Tomchenko17}.

Using the fact that $\delta k $ vanishes slower than $c$ in the limit $c\rightarrow 0^+$ the equation (\ref{GauSol}) reduces to
\be
 \lim_{ \, \, \, c\rightarrow 0^+}  \! \! \!  \Psi(\{x\},\{q\}) = \! \!  \!  \! \! \! \! \! \! \sum_{\sigma \in \mathcal{S}_{N},\{\epsilon\}_N} \! \! \! \! \! \! \! \! \epsilon_1  \! \cdot \! \cdot \!  \cdot \epsilon_N\exp\left[ i \sum_{s=1}^{N} q_{\sigma(s)} x_s\right]. 
\label{app4reduction}
\ee \\
\begin{widetext}
Now we want to show the following identity
\bea
\displaystyle{\sum_{\sigma \in \mathcal{S}_{N}}\sum_{\{\epsilon\}_N}  \epsilon_1   \cdot  \cdot   \cdot \epsilon_N\exp\left[ i \sum_{s=1}^{N} \epsilon_{\sigma(s)} k_{\sigma(s)} x_s\right] } 
 = \sum_{\sigma \in \mathcal{S}_{N}} \prod_{s=1}^{N}\left( \mathrm{e}^{i k_{\sigma(s)} x_s} -\mathrm{e}^{-i k_{\sigma(s)} x_s} \right).
\label{app5id}
\eea
It is clear that it is enough to show the identity for a single arbitrary permutation $\sigma$, 
\bea
\sum_{\, \, \, \{\epsilon\}_{N}} \epsilon_1 \! \cdot \! \cdot \! \cdot \epsilon_{N}\exp\left[ i \sum_{s=1}^{N} \epsilon_{\sigma(s)} k_{\sigma(s)} x_s\right] 
= \prod_{s=1}^{N}\left( \mathrm{e}^{i k_{\sigma(s)} x_s} -\mathrm{e}^{-i k_{\sigma(s)} x_s} \right).
\label{appeneq}
\eea
In general, we can change positions of epsilons (or equivalently permute indices) in the following way $\epsilon_{1}\epsilon_{2} \cdot \! \cdot \! \cdot \epsilon_{N} = \epsilon_{3}\epsilon_{1} \cdot \! \cdot\! \cdot \epsilon_{N} \cdot \! \cdot \! \cdot \epsilon_{5} = \epsilon_{\sigma(1)}\epsilon_{\sigma(2)} \cdot \! \cdot \! \cdot \epsilon_{\sigma(N)}$. Now, the equality (\ref{appeneq}) can be shown using the fact that 
$$\displaystyle{ \sum_{\, \, \, \{\epsilon\}_{N}} \epsilon_1 \! \cdot \! \cdot \! \cdot \epsilon_{N}\exp\left[ i \sum_{s=1}^{N} \epsilon_{\sigma(s)} k_{\sigma(s)} x_s\right] 
}
=
\displaystyle{ 
 \left( \sum_{\epsilon_{\sigma(1)}} \epsilon_{\sigma(1)} \mathrm{e}^{i\epsilon_{\sigma(1)} k_{\sigma(1)}x_1} \right) \left( \sum_{\epsilon_{\sigma(2)}} \epsilon_{\sigma(2)} \mathrm{e}^{i\epsilon_{\sigma(2)} k_{\sigma(2)}x_2} \right)\cdot \! \cdot \! \cdot \left( \sum_{\epsilon_{\sigma(N)}} \epsilon_{\sigma(N)} \mathrm{e}^{i\epsilon_{\sigma(N)} k_{\sigma(N)}x_N} \right) },$$

\end{widetext}
 which ends the proof because every single term of the product,
\bea 
\sum_{\epsilon_{\sigma(r)}} \epsilon_{\sigma(r)} \mathrm{e}^{i\epsilon_{\sigma(r)} k_{\sigma(r)}x_r} &=& \mathrm{e}^{i k_{\sigma(r)}x_r} -\mathrm{e}^{-i k_{\sigma(r)}x_r}
\cr &=&
2i\sin\left(k_{\sigma(r)}x_r\right).
\eea

Hence, in the considered limit, the equation (\ref{GauSol}) can be rewritten in terms of sine functions
\be
\lim_{ \, \, \, c\rightarrow 0^+} \Psi(\{x\},\{k\})=\left(2i \right)^N \sum_{\sigma\in \mathcal{S}_N}\prod_{s=1}^{n}\sin \left(k_{\sigma(s)} x_s \right).
\label{AppGauReduced} 
\ee


\begin{thebibliography}{99}

\bibitem{KivsharOpticalSol}
Y. S. Kivshar and G. P. Agrawal, {\it Optical Solitons}, Academic Press, An
imprint of Elsevier Science, San Diego, California, 2003.

\bibitem{pethicksmith}
C. Pethick and H. Smith, {\it Bose-Eistein condensation in dilute gases} (Cambridge University Press, Cambridge, England, 2002).

\bibitem{burger1999}
S. Burger, K. Bongs, S. Dettmer, W. Ertmer, K. Sengstock, A. Sanpera, G. V. Shlyapnikov, and M. Lewenstein,
Phys. Rev. Lett. {\bf 83}, 5198 (1999).

\bibitem{denschlag2000}
J. Denschlag, J. E. Simsarian, D. L. Feder, Charles W. Clark, L. A. Collins, J. Cubizolles, L. Deng, E. W. Hagley, K. Helmerson, W. P. Reinhardt, S. L. Rolston, B. I. Schneider, and W. D. Phillips, Science {\bf 287}, 97 (2000).

\bibitem{strecker2002}
K. E. Strecker, G. B. Partridge, A. G. Truscott, and R. G. Hulet,
Nature {\bf 417}, 150 (2002).

\bibitem{khaykovich2002}
L. Khaykovich, F. Schreck, G. Ferrari, T. Bourdel, J. Cubizolles, L. D. Carr, Y. Castin, and C. Salomon,
Science {\bf 296}, 1290 (2002).

\bibitem{becker}
C. Becker, S. Stellmer, P. Soltan-Panahi, S. D\"orscher, M. Baumert, E.-M. Richter, J. Kronj\"ager, K. Bongs, and K. Sengstock, Nature Physics {\bf 4}, 496 (2008).



\bibitem{Stellmer2008}
S. Stellmer, C. Becker, P. Soltan-Panahi, E.-M. Richter, S. D\"orscher, M. Baumert, J. Kronj\"ager, K. Bongs, and K. Sengstock, Phys. Rev. Lett. {\bf 101}, 120406 (2008).

\bibitem{Weller2008}
A. Weller, J. P. Ronzheimer, C. Gross, J. Esteve, M. K. Oberthaler, D. J. Frantzeskakis, G. Theocharis, and P. G. Kevrekidis, Phys. Rev. Lett. {\bf 101}, 130401 (2008).

\bibitem{Theocharis2010}
G. Theocharis, A. Weller, J. P. Ronzheimer, C. Gross, M. K. Oberthaler, P. G. Kevrekidis, and D. J. Frantzeskakis, Phys. Rev. A {\bf 81}, 063604 (2010).



 


\bibitem{dziarmaga2004}
J. Dziarmaga, Phys. Rev A. {\bf 70}, 063616 (2004).

\bibitem{Mishmash2009_1}
R. V. Mishmash and L. D. Carr, Phys. Rev. Lett. {\bf 103}, 140403 (2009). 

\bibitem{Mishmash2009_2}
R. V. Mishmash, I. Danshita, Charles W. Clark, and L. D. Carr, Phys. Rev. A {\bf 80}, 053612 (2009).

\bibitem{Dziarmaga2010}
J. Dziarmaga, P. Deuar, and K. Sacha, Phys. Rev. Lett. {\bf 105}, 018903 (2010).

\bibitem{Mishmash2010}
R. V. Mishmash and L. D. Carr,
Phys. Rev. Lett. {\bf 105}, 018904 (2010).

\bibitem{delande2014}
D. Delande and K. Sacha, Phys. Rev. Lett. {\bf 112}, 040402 (2014).

\bibitem{kronke15} 
S. Kr\"onke and P. Schmelcher, Phys. Rev. A {\bf 91}, 053614 (2015). 

\bibitem{Hans2015}
G. C. Katsimiga, G. M. Koutentakis, S. I. Mistakidis, P.~G. Kevrekidis, and P. Schmelcher, arXiv:1612.09151.

\bibitem{Yefsah}
T. Yefsah, A. T. Sommer, M. J. H. Ku, L. W. Cheuk, W. Ji, W. S. Bakr, and M. Zwierlein, Nature {\bf 499}, 426 (2013).




\bibitem{roberts2000}
J. L. Roberts, N. R. Claussen, S.L. Cornish, and C. E. Wieman,
Phys. Rev. Lett. {\bf 85}, 728 (2000).

\bibitem{plodzien2012}
M. P\l{}odzie\'n and K. Sacha, Phys. Rev. A {\bf 86}, 033617 (2012).

\bibitem{Lai89}
Y. Lai and H. A. Haus, Phys. Rev. A 40, 844 (1989).

\bibitem{Lai89a}
Y. Lai and H. A. Haus, Phys. Rev. A 40, 854 (1989).

\bibitem{castinleshouches}
Y. Castin, in Les Houches Session LXXII, {\it Coherent atomic matter waves 1999}, edited by R. Kaiser, C. Westbrook and F. David, (Springer-Verlag Berlin Heilderberg New York 2001).

\bibitem{Weiss09}
C. Weiss and Y. Castin, Phys. Rev. Lett. {\bf 102}, 010403 (2009).

\bibitem{delande2013}
D. Delande, K. Sacha, M. P\l{}odzie\'n, S. K. Avazbaev, and J. Zakrzewski, New J. Phys. {\bf 15}, 045021 (2013).

\bibitem{corney97}
J. F. Corney, P. D. Drummond, and A. Liebman, Opt. Commun.
{\bf 140}, 211 (1997).

\bibitem{corney01}
J. F. Corney and P. D. Drummond, J. Opt. Soc. Am. B {\bf 18}, 153 (2001).

\bibitem{martin2010b}
A. D. Martin and J. Ruostekoski, New J. Phys. {\bf 12}, 055018 (2010).



\bibitem{Boisse2017}
A. Boiss\'e, G. Berthet, L. Fouch\'e, G. Salomon, S. Aspect, S. Lepoutre, and T. Bourdel, arXiv:1701.00414.

\bibitem{Bethe31}
H. Bethe, Z. Physik {\bf 71}, 205 (1931). 


\bibitem{Korepin93}
V. E. Korepin, N. M. Bogoliubov, and A. G. Izergin, {\it Quantum Inverse Scattering Method and Correlation Functions} (Cambridge University Press, Cambridge, 1993).

\bibitem{Gaudin}
M. Gaudin, {\it The Bethe wavefunction}, Cambridge University Press, 2014.

\bibitem{Oelkers2006}
N. Oelkers, M. T. Batchelor, M. Bortz, and X. W. Guan, J. Phys. A: Math. Gen. {\bf 39}, 1073 (2006).

\bibitem{Lieb63}
E. H. Lieb and W. Liniger, Phys. Rev. {\bf 130}, 1605 (1963).

\bibitem{Lieb63a}
E. H. Lieb, Phys. Rev. {\bf 130}, 1616 (1963).


\bibitem{kulish76}
P. P. Kulish, S. V. Manakov, and L. D. Faddeev, Theor. Math. Phys. {\bf 28}, 615 (1976).

\bibitem{ishikawa80}
M. Ishikawa and H. Takayama, J. Phys. Soc. Jpn. {\bf 49}, 1242 (1980).

\bibitem{komineas02}
S. Komineas and N. Papanicolaou, Phys. Rev. Lett. {\bf 89}, 070402 (2002). 

\bibitem{jackson02}
A. D. Jackson and G. M. Kavoulakis, Phys. Rev. Lett. {\bf 89}, 070403 (2002).

\bibitem{kanamoto08}
R. Kanamoto, L. D. Carr, and M. Ueda, Phys. Rev. Lett. {\bf 100}, 060401 (2008).

\bibitem{kanamoto10}
R. Kanamoto, L. D. Carr, and M. Ueda, Phys. Rev. A {\bf 81}, 023625 (2010); Erratum Phys. Rev. A {\bf 81}, 049903(E) (2010).

\bibitem{karpiuk12}
T. Karpiuk, P. Deuar, P. Bienias, E. Witkowska, K. Paw\l{}owski, M. Gajda, K. Rz\c{a}\.zewski, and M. Brewczyk, Phys. Rev. Lett. {\bf 109}, 205302 (2012).

\bibitem{karpiuk15}
T. Karpiuk, T. Sowi\'nski, M. Gajda, K. Rz\c{a}\.zewski, and M. Brewczyk, Phys. Rev. A {\bf 91}, 013621 (2015).

\bibitem{sato12}
J. Sato, R. Kanamoto, E. Kaminishi, and T. Deguchi, Phys. Rev. Lett. {\bf 108}, 110401 (2012).

\bibitem{sato12a}
J. Sato, R. Kanamoto, E. Kaminishi, and T. Deguchi, preprint arXiv:1204.3960.

\bibitem{sato16}
J. Sato, R. Kanamoto, E. Kaminishi, and T. Deguchi, New J. Phys. {\bf 18}, 075008 (2016).

\bibitem{Gawryluk2017}
K. Gawryluk, M. Brewczyk, and K. Rz\c{a}\.zewski,
Phys. Rev. A {\bf 95}, 043612 (2017).

\bibitem{Syrwid2015}
A. Syrwid and K. Sacha, Phys. Rev. A {\bf 92}, 032110 (2015).

\bibitem{Syrwid2016}
A. Syrwid, M. Brewczyk, M. Gajda, and K. Sacha, Phys. Rev. {\bf A} 94, 023623 (2016).

\bibitem{Gaudin71}
M Gaudin, Phys. Rev. A {\bf 4}, 386 (1971).

\bibitem{Batchelor05}
M. T. Batchelor, X. W. Guan, N. Oelkers, C. Lee, J. Phys. A {\bf 38}, 7787-7806 (2005).

\bibitem{Tomchenko17}
M. Tomchenko, J. Phys. A: Math. Theor. {\bf 50}, 055203 (2017).

\bibitem{McGuire}
J. B. McGuire, J. Math. Phys. {\bf 5}, 622 (1964).


\bibitem{javanainen96}
J. Javanainen and S. M. Yoo, Phys. Rev. Lett. {\bf 76}, 161 (1996).

\bibitem{dziarmaga03}
J. Dziarmaga, Z. P. Karkuszewski, and K. Sacha, J. Phys. B, {\bf 36}, 1217 (2003).

\bibitem{dziarmaga06}
J. Dziarmaga and K. Sacha, J. Phys. B, {\bf 39}, 57 (2006).

\bibitem{Dagnino09}
D. Dagnino, N. Barber\'an, and M. Lewenstein, Phys. Rev. A {\bf 80}, 053611 (2009).

\bibitem{Kasevich20016}
K. Sakmann and M. Kasevich, Nature Physics {\bf 12}, 451-454 (2016).



\bibitem{Metropolis1953}
M. Metropolis, A. W. Rosenbluth, M. N. Rosenbluth, A. H. Teller, and E. Teller,
J. Chem. Phys. {\bf 21}, 1087 (1953).


\bibitem{Gajda_PauliCrystal}
M. Gajda, J. Mostowski, T. Sowi\'n{}ski, and M. Za\l{}uska-Kotur, EPL {\bf 115}, 20012 (2016).

\bibitem{Carr_HWSoliton}
L. D. Carr, Charles W. Clark, and W. P. Reinhardt, Phys. Rev. {\bf A} 62, 063610  (2000).

\bibitem{girardeau1960}
M. Girardeau, J. Math. Phys. (NY) {\bf 1}, 516 (1960).



\bibitem{Paredes2004}
B. Paredes, A. Widera1, V. Murg, O. Mandel, S. F\"olling, I. Cirac, G. V. Shlyapnikov, T. W. H\"ansch, and I. Bloch, Nature {\bf 429}, 277 (2004).














\end{thebibliography}
\end{document}